\documentclass[aps,pre,preprint,showpacs,showkeys,groupedaddress]{revtex4}
\newcommand{\W}{14cm}
\newcommand{\Ws}{12cm}

\usepackage{graphicx}
\usepackage{amsmath}
\begin{document}
\title{Separation of suspended particles in microfluidic systems by directional-locking in periodic fields}
\author{John Herrmann}
\author{Michael Karweit}
\author{German Drazer}
\email{drazer@mailaps.org}
\affiliation{Department of Chemical \& Biomolecular Engineering, Johns Hopkins University, Baltimore, Maryland, 21218}
\date{\today}

\begin{abstract}
We investigate the transport and separation of overdamped particles under the action of a uniform external force
in a two-dimensional periodic energy landscape. Exact results are obtained for the deterministic transport 
in a square lattice of parabolic, repulsive centers that correspond to a piecewise-continuous linear-force model. 
The trajectories are periodic and commensurate with the obstacle lattice and exhibit phase-locking behavior in that 
the particle moves at the same average migration angle for a range of orientation of the 
external force. The migration angle as a function of the orientation of the external force has a Devil's staircase 
structure. The first transition in the migration angle was analyzed in terms of a
Poincare map, showing that it corresponds to a tangent bifurcation. Numerical results show that
the limiting behavior for impenetrable obstacles is equivalent to the high Peclet number limit
in the case of transport of particles in a periodic pattern of solid obstacles. 
Finally, we show how separation occurs in these systems depending on the properties of the particles.

\end{abstract}
\maketitle

\section{Introduction}
\label{intro}

The transport of suspended particles in a two-dimensional (2D) periodic periodic lattice of obstacles
has recently attracted significant attention as a means of separating particle species in microfluidic devices. 
A driving force induces particles to move through the system. Depending on the properties of the particles and 
the direction of the driving force $\theta$ with respect to the lattice orientation, the particles move through
the lattice in commensurate phase-locked trajectories. That is, for a range of driving-force angles single species of
particles will have an average trajectory in one of the lattice directions $[p,q]$, whose angle is
given by $\alpha=\arctan(q/p)$. Further, that average trajectory angle will remain fixed over a range of driving-force angles. 
When
the driving-force angle is outside that range, the particle trajectories will jump to a new fixed angle. 
The relation between the direction of motion $\alpha$ and the angle of the driving force $\theta$ 
depends on the properties of the particles. Consequently, the system has the potential to 
separate different species of particles, with the advantage that 
different particles migrate at different angles ({\it vector chromatography}), allowing for continuous
fractionation \cite{DorfmanB01,DorfmanB02}.

These microfluidic systems can be broadly classified into two groups, depending on the nature of the 
energy landscape that is experienced by the suspended particles. In one group, the particles move through a 
periodic array of solid obstacles, such as in the separation devices based on {\it deterministic 
hydrodynamics} pioneered by Austin's group \cite{HuangCAS04,MortonLITSCA08,MortonLITSCA08b,BeechBT08}. 
In terms of the corresponding energy landscape, the array of obstacles can be considered as a periodic 
array of {\it hard-core} repulsive potentials. In the second class of systems, particles move through 
smooth potential landscapes, with {\it soft} interactions between the particles and an external field. 
Examples include the optical fractionation methods pioneered by Grier's group, where colloidal particles 
are transported through an array of holographic 
optical tweezers \cite{KordaTG02,LadavacLKG04,AjayG04,PeltonLG04,RoichmanRWG07}. 
 
Phase-locking behavior is common to transport through
periodic structures in many systems \cite{ReichhardtF99,MarconiCBPDM00,ReichhardtOH02,ReichhardtRH04,ReichhardtR04} 
as well as to non-linear dynamical systems in general \cite{Strogatz94,Ott02}.  
In the transport of particles in periodic systems, \textcite{LacastaSRL05}
studied a 2D periodic arrangement of wells (or traps) in a square lattice
by means of numerical simulations of the corresponding Langevin equation.
Numerical results showed the presence of periodic trajectories and lateral migration, in that the
particles moved, on average, at an angle $\alpha$ different from the orientation angle of the driving force $\theta$.
They also showed the presence of phase locking in the $\alpha$ vs. $\theta$ curve, with clear plateaus at
large Peclet numbers. In fact, the authors identify the observed migration as a deterministic
phenomenon \cite{SanchoKLL05}. 

In a separate study, \textcite{LacastaKSL06} investigated periodic landscapes that
present either repulsive (obstacles) or attractive (traps) centers located on a square lattice. 
In both repulsive and attractive cases, they observe similar phase locking behavior, with the corresponding 
plateaus in the $\alpha$ vs. $\theta$ curve becoming evident at large Peclet numbers 
(Large Peclet numbers corresponds to low temperatures in Refs. \cite{LacastaSRL05} and \cite{LacastaKSL06}). 

On the other hand, separable potentials, which can be written as a sum of periodic fields in each of the two principal directions
of a square lattice \cite{SanchoKLL05}, do not exhibit the same complex behavior, 
as clearly shown by \textcite{PeltonLG04}.
Experimental work using optical tweezers has also demonstrated the presence of periodic 
trajectories and locked-in states \cite{PeltonLG04,KordaTG02}. However, these experiments displayed 
a different type of locking in the presence of Brownian motion, in which the trajectories become 
commensurate only in a statistical sense and only some of the locked-in states
are centered on commensurate directions (A hopping model for the observed dynamics is proposed by \textcite{AjayG04}). 

Conversely, and in agreement with the discussed numerical results at high Peclet 
numbers, recent simulations and experiments on the motion of non-Brownian spherical particles through periodic 
arrays of obstacles clearly show that deterministic systems exhibit analogous phase-locking dynamics and periodic trajectories 
into commensurate lattice directions \cite{FrechetteD09}. 

Phase-locking dynamics has also been observed in more complex transport systems, 
including the motion of an overdamped particle through a periodic potential under time-periodic driving forces \cite{GuantesM03,ReichhardtRH04,ReichhardtOH02} and the dynamics of a particle driven through a deformable colloidal lattice \cite{ReichhardtR04}.

In contrast to a significant number of experiments and simulations, there are few studies that present analytical results for the 
transport of particles in irreducible two-dimensional systems. \textcite{PeltonLG04} studied the overdamped motion of a single 
particle in a periodic landscape. In the 1D case (linear fringes) they showed the existence of locked trajectories for 
driving forces that are not strong enough to drive the particles over the energy barrier. They also showed that the 
behavior in separable (reducible) 2D potentials is analogous to the locking observed in 1D systems. 
\textcite{GleesonSLL06} studied irreducible 2D periodic potentials and derived an iterative method for 
calculating the average velocity of the particles
in inverse powers of the external force and particle's diffusivity. 

In this work, we are interested in the deterministic limit (small diffusivity)
for relatively small forces, {\it i. e.}, driving forces for which the effect of the potential landscape is not negligible.
Therefore, we consider both the deterministic evolution as well as the high Peclet number limit for finite forces.
In particular, we consider a quadratic, continuous model for the 2D periodic potential
that corresponds to a piecewise-continuous, linear-force model. This simple model offers the interesting
combination of an irreducible but solvable potential that captures the 
non-trivial transport phenomena exhibited by 2D periodic systems, including directional locking 
and the universal behavior of dynamical systems near a bifurcation point. 

\section{Transport of Colloidal particles through periodic landscapes}
\label{transport}

\subsection{Equation of Motion: High Friction Limit} 
\label{subsec:langevin}

The equation of motion for a Brownian particle traversing a periodic force field is the Langevin 
equation \cite{McQuarrie00}, which in the high-friction limit takes the form \cite{Risken}: 
\begin{equation}
\label{langevin}
\gamma \frac{d {\bf x}}{dt} = {\bf F}({\bf x})+{\bf F}_0({\bf x})+{\bf \xi}(t),
\end{equation}
where ${\bf F}({\bf x})$ is the periodic force field, ${\bf F}_0$ is an external driving force and ${\bf \xi}(t)$ 
is the Langevin force describing the fluctuating force exerted by the fluid on the colloidal particle. 
The friction constant $\gamma$ is given by $6\pi\mu a$, where $a$ is the radius of the colloid and $\mu$ 
is the viscosity of the fluid. The Langevin force is represented by a Gaussian distribution with
zero mean, $\langle \mathbf{\xi}(t)\rangle=0$, and $\delta$ correlation, 
$\langle \xi_i(t)\xi_j(s) \rangle=2 \gamma kT \delta(t-s)\delta_{ij}$. Here, we shall focus on a 
spatially uniform external force, ${\bf F}_0({\bf x})\equiv {\bf F}_0$ and, for convenience, we
choose a coordinate system with the $x$ axis oriented along the direction of the force,
${\bf F}_0({\bf x})= F_0 \, {\bf e}_x$ (see figure \ref{quadratic}). 

\subsection{Periodic Potential: Piecewise-linear model} 
\label{subsec:potential}
\begin{figure}
\centering
\includegraphics*[width=\W]{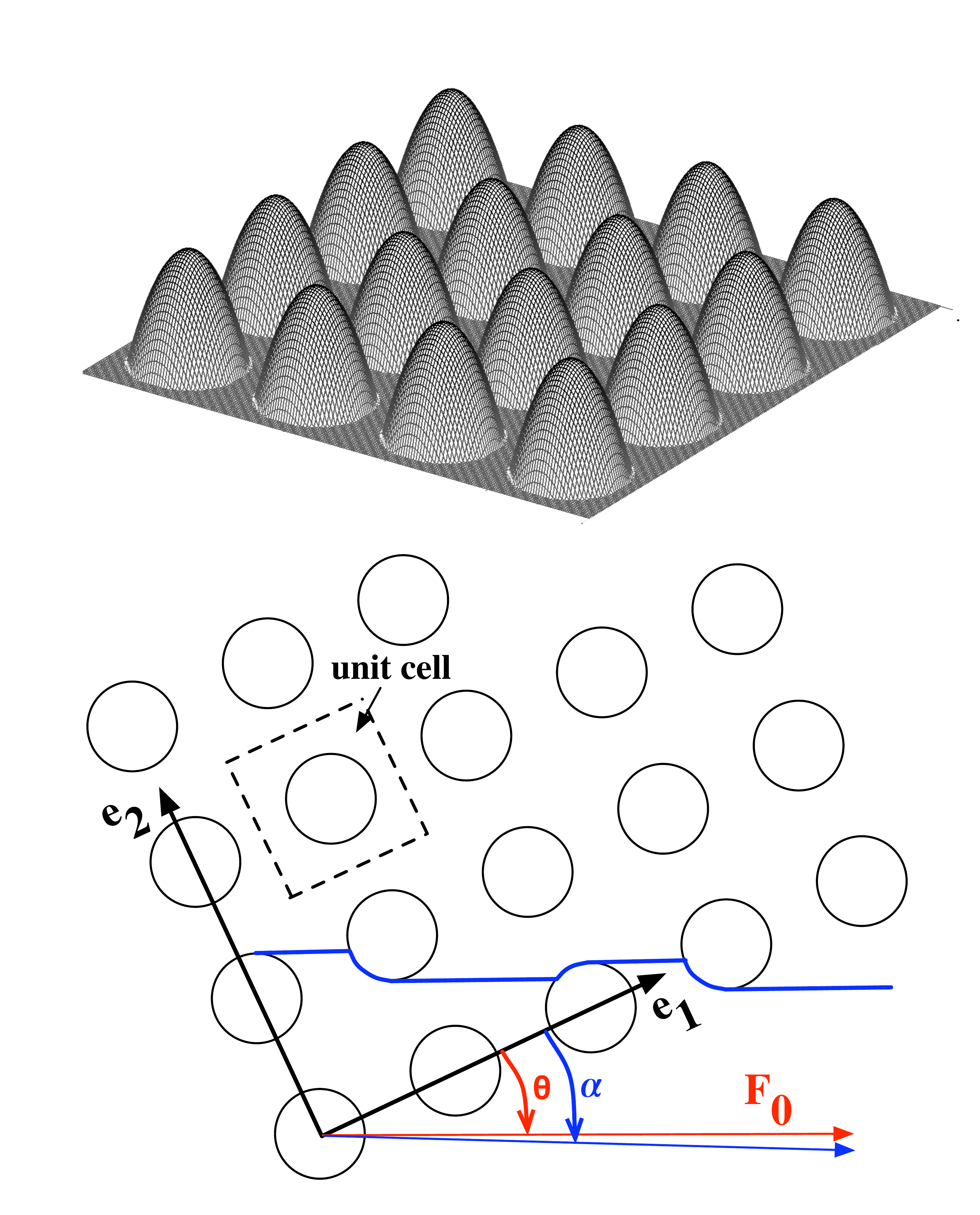}
\caption[]{(Top) Potential landscape with quadratic, repulsive centers on a square lattice.
(Bottom) Schematic view of the system, with circles representing the repulsive centers. ${\bf e}_1$ and
${\bf e}_2$ are the principal vector of the square lattice. A unit cell is represented by the dashed line.
The $F_0$ is oriented along the $x$ axis, and $\theta$ is the angle between the force and ${\bf e}_1$ 
(in this case is $\theta=25^\circ$). The solid line shows a locked trajectory that moves in the [2,1]
lattice direction, indicated with a solid arrow at the bottom of the system. $\alpha$ is 
the angle between this asymptotic direction and ${\bf e}_1$: $\alpha=\arctan(1/2)=26.56^\circ$.}
\label{quadratic}
\end{figure}

We consider the case in which the periodic force can be derived from a potential field, ${\bf F}({\bf x})=-\nabla V({\bf x})$.
Here, we shall model the two-dimensional, periodic landscape as a piecewise-smooth potential that is composed
of repulsive centers of size $R$ ({\em obstacles}) located in a square lattice with lattice spacing $L>R$. 
Specifically, we consider the periodic landscape shown in figure \ref{quadratic} which, in 
the unit cell, is given by,
\begin{equation}
V(x,y)=\left\{ 
\begin{array}{lr}
-\frac{F_{max}}{2R}\left(x^2+y^2-R^2\right) & \qquad r \leq R \\
0 & \qquad r > R,
\end{array}
\right.
\end{equation}
where the center of the coordinate system coincides with the center of the cell, $r$ is the radial
position, $r^2=x^2+y^2$, and $F_{max}$ gives the magnitude of repulsive centers. Since we have 
chosen to align the $x$ axis with the external force the obstacle lattice will be, in general,
rotated with respect to the coordinate system. 
We shall refer to the rotation angle between the $x$ axis of the coordinate system
and the principal axis of the square lattice ${\bf e}_1$ as the forcing angle $\theta$ (see figure \ref{quadratic}). 
We shall also non-dimensionalize our variables using $u_c=F_{max}/\gamma$ as the characteristic velocity, 
$F_{max}$ as the characteristic force, and $R$ as the characteristic length. The new variables become 
$\dot {\bf x}' = \dot{\bf x} \gamma/F_{max}$; ${\bf x}' = {\bf x}/R$; $r' = r/R$. The boundary of the repulsive centers
is at $r'=1$, and we define the relative separation between the repulsive centers 
as $\ell=L/R$. For simplicity, we do not use the primes to refer to the non-dimensional variables in what follows.

\section{Deterministic Transport: Exact Solutions}

The trajectories in the deterministic limit are obtained from Eqs. \ref{langevin} by neglecting the effect of 
thermal fluctuations. Thus, outside the quadratic regions, the particles follow a straight line that is parallel to
the $x$ axis. On the other hand, inside the parabolic regions, the equations of motion become,
\begin{eqnarray}
\label{diffeq}
\dot x &=& -\frac{\partial V}{\partial x} + f = x + f, \qquad r<1,\\ \nonumber
\dot y &=& -\frac{\partial V}{\partial y} = y, \qquad r<1,
\end{eqnarray}
with $f=F_0/F_{max}$ the ratio of the driving to the repulsive force.

We can include the external force in a modified potential field as $V_m(x,y)=V(x,y)+x\;f$.
This modified potential is also an inverted parabola, but with its center 
shifted to ${\bf x}_0=(-f,0)$. It is clear then that the family of curves perpendicular to the equipotential lines, 
{\it i. e.}, the particle trajectories inside the circle, are also straight lines, 
with center at ${\bf x}_0$. The same result can be obtained from direct integration of Eqs. \ref{diffeq}. 

\begin{figure}
\includegraphics*[width=10cm]{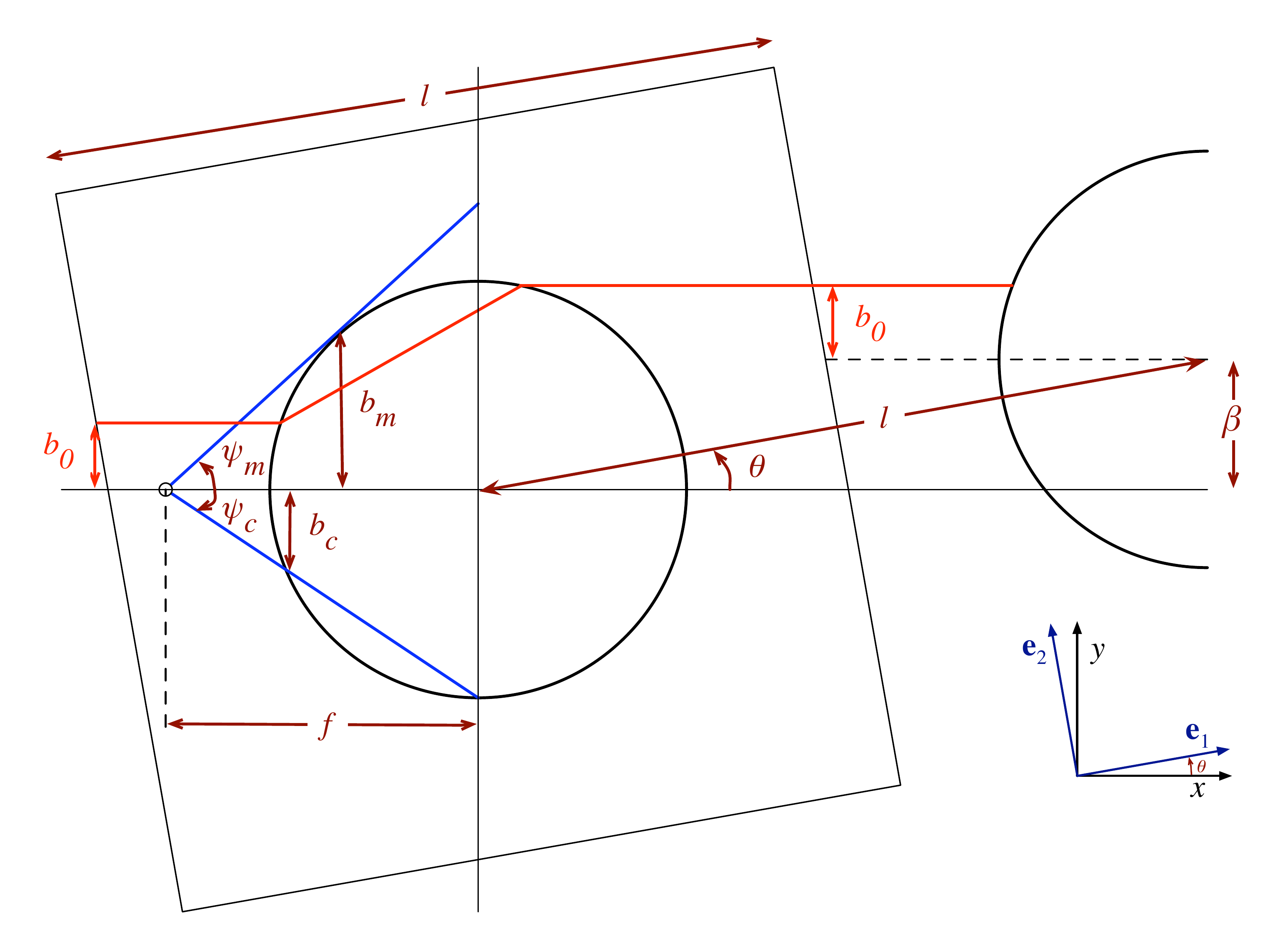}
\includegraphics*[width=10cm]{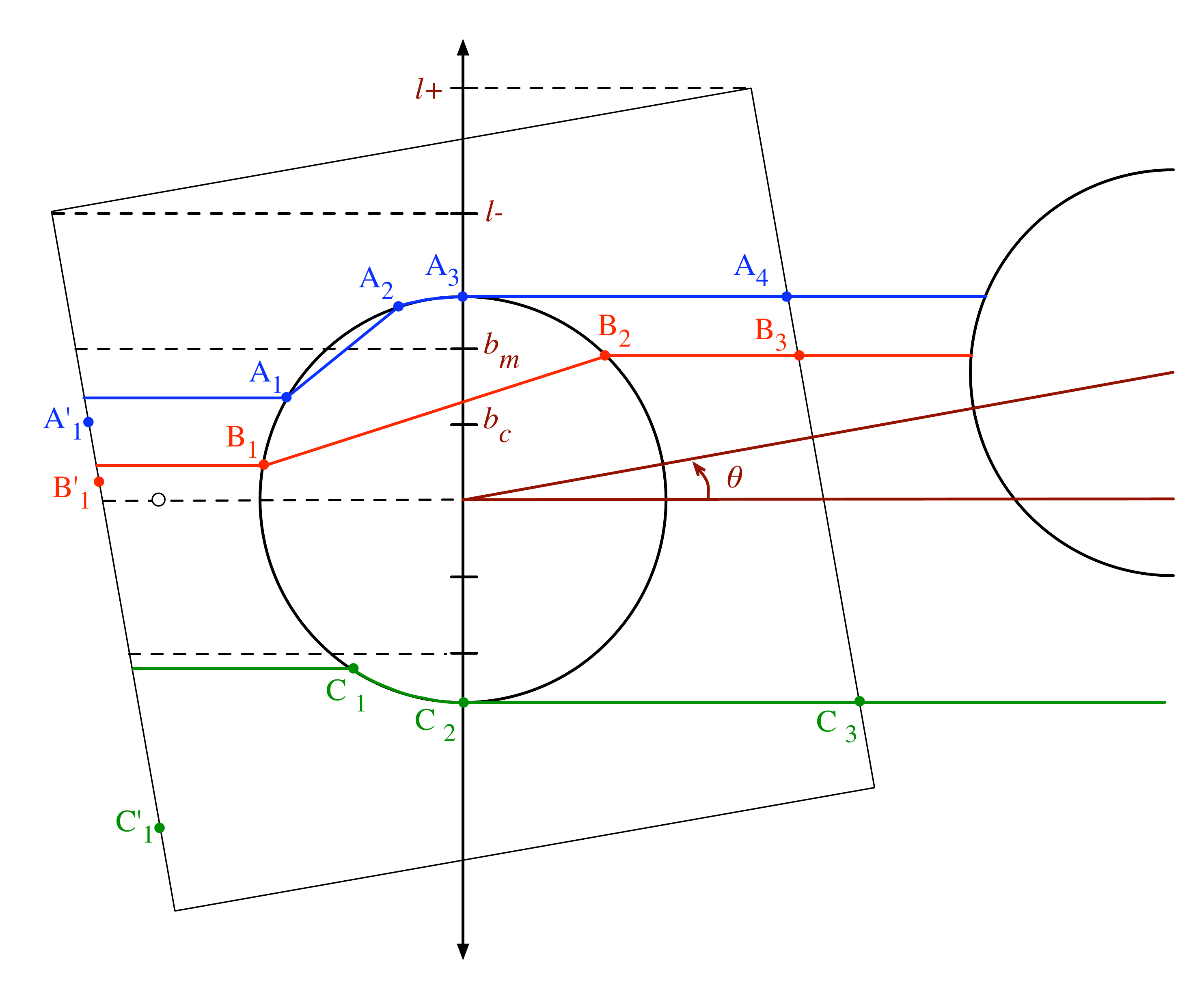}
\caption[]{a) Schematic view of a unit cell for a rotation angle $\theta$. $b_0$ is the incoming impact 
parameter for the trajectory shown with the solid line. We also show the incoming impact parameter for 
the next collision and the corresponding shift by $\beta$. 
$\ell$ is the ratio between center size $R$ and center-to-center distance $L$.
The center for the collision trajectories is $(-f,0)$. 
The critical impact parameters are $b_c$, which corresponds to a collision trajectory 
ending at $(0,\pm1)$, and $b_m$, which corresponds to a collision trajectory tangent to the parabolic 
center. b) Schematic view of the three types of trajectories: 
Trajectory $A$ for incoming parameters $b_c<b_o<b_m$; $B$ for $0<b_0<b_c$; and $C$ for $b_m<b_0<1$.
$\ell_+$ is the maximum value of the incoming parameter for a given forcing direction $\theta$. 
The region between $\ell_+$ and $\ell_-$ corresponds to the trajectories crossing the top 
(bottom) of the unit cell.
}
\label{definitions}
\end{figure}

When $f>1$ the $x$ component of the velocity is always positive and all the trajectories are unbounded. 
Then, we can define the migration angle $\alpha$ of the particles as the asymptotic angle that the trajectory 
of the particles forms with the principal axis of the lattice ${\bf e}_1$, {\it i. e.}, $\alpha=\theta$ means 
that the particles move colinearly with the external force; 
$\alpha=0$ means that the particle moves parallel to the principal direction of the square lattice 
(see figs. \ref{quadratic} and \ref{definitions}). 

Let us note that we are in the deterministic and high-friction limits, and therefore two independent trajectories cannot
cross each other in real space. In addition, we consider two-dimensional trajectories in the plane, and thus 
for a given value of the forcing angle $\theta$ all the trajectories must have the same migration angle $\alpha$. 
Therefore, we can obtain a single valued $\alpha$ vs. $\theta$ curve by determining the angle of a single trajectory 
as a function of the forcing angle.

\subsection{Analytical solutions in the unit cell}

In general, a global trajectory can be segmented into a number of successive {\it collisions} between the colloidal
particle and the repulsive centers. In addition, we can classify each trajectory passing through a unit cell in terms of 
its incoming impact parameter $b_0$ which we define in the present coordinate system as the $y$ coordinate of the particle 
when it enters the unit cell (see fig. \ref{definitions}a). For $b_0>1$ the trajectories are straight lines parallel 
to the $x$ axis and do not interact with the obstacle at the center of the cell. For $0<b_0<1$, on the 
other hand, we have three different cases. For impact parameters larger than a maximum value, 
\begin{equation}
b_m=\frac{1}{f}\sqrt{f^2-1},
\end{equation}
it can be shown from Eqs. \ref{diffeq} that the radial component of the velocity is negative outside the circle and positive 
inside it:
\begin{equation}
\dot r =\left\{ 
\begin{array}{ll}
xf/r < 0& \qquad {\rm for} \qquad r>1 \\
xf/r+r > 0 & \qquad {\rm for} \qquad r<1 \;\; \& \;\; 0<b_0<b_m.
\end{array}
\right.
\end{equation}
In our piecewise approximation, this means that the particle will move around the circle with $r=1$, 
until it separates at the top of the circle and then follows a straight line parallel to the $x$ axis. 
A trajectory of this type is shown in  fig. \ref{definitions}b: the incoming particle enters the circle at the point $C_1$ 
with impact parameter $b_0<-b_m$, and therefore leaves the circle at $C_2=(0,-1)$.
The maximum value of the impact parameter, $b_m$, corresponds to the point on the circumference for which its tangent passes through 
the center of the effective potential, $(-f,0)$, as shown in fig. \ref{definitions}a.

For impact parameters smaller than $b_m$ there is a critical value,
\begin{equation}
b_c= \left(\frac{f^2-1}{f^2+1}\right),
\end{equation}
such that, for $b_c<b_0<b_m$ the particle enters the repulsive centers, but the separation 
from the circle still occurs at the top (or bottom for negative values of the impact parameter, see below). 
Such a characteristic trajectory is shown in fig. \ref{definitions}b: the particle enters the circle at 
the point $A_1$ and follows a straight line with center at $(-f,0)$. 
Then, the particle reaches the point $A_2$ on the circumference of the circle, which belongs to the region above $b_m$. Thereafter,
the trajectory becomes identical to those described for $b_0>b_m$, with the particle separating from the circle at the top (point $A_3$).
We will refer to collisions with incoming parameter larger than $b_c$ as {\it irreversible} in that, independent of the exact value of the impact parameter, all the incoming trajectories collapse into a single outgoing trajectory, with outgoing impact parameter $b_f=1$. 

Finally, for impact parameters smaller than the critical value, $0<b_0<b_c$, the particle enters the circle 
(e. g. point $B_1$ in figure \ref{definitions}b), 
moves in a straight line, and leaves the circle on the positive side of the $x$ axis (point $B_2$). The outgoing impact parameter 
(or $y$ coordinate) in this case is given by,
\begin{equation}
b_f=\frac{\left( f^2-1\right) \, b_0}{\left( f^2 + 1 \right)-2 f \sqrt{1-b_0^2}}
\end{equation}

\subsection{Poincare Map and Saddle-Point Bifurcation}

In figure \ref{definitions}b we showed the different types of trajectories that a particle follows depending on the impact 
parameter. For each one of these trajectories, we also show the impact parameter that the particle will have in its next collision
in a neighboring unit cell (see points $A'_1$, $B'_1$, and $C'_1$). 
If the outgoing impact parameter $b_f \equiv b_f(b_0)$ is defined as the $y$ coordinate of the
point where the particle leaves the cell,  then the next incoming parameter will be $b'_0=b_f-\beta$, with $\beta=\ell \sin(\theta)$,
as shown in figure \ref{definitions}a. 
Therefore, for positive values of the incoming impact parameter, the 
impact parameter for the next collision is given by:
\begin{equation}
\label{map}
b'_0 = \left\{ 
\begin{array}{ll}
\frac{\left( f^2-1\right) \, b_0}{\left( f^2 + 1 \right)-2 f \sqrt{1-b_0^2}}-\beta & \qquad 0< b_0 < b_c \\
1-\beta& \qquad b_c<b_0<1 \\
b_0-\beta& \qquad 1<b_0<\ell_+
\end{array}
\right.
\end{equation}
where $\ell_+=(\ell/2) (\cos(\theta)+\sin(\theta))$ is the maximum possible value of the incoming impact parameter (see fig. 
\ref{definitions}b). The symmetric conditions apply to negative values of the impact parameter, 
with the only difference being that those trajectories leaving at the bottom of the cell come into the next cell from the top, that is:
\begin{equation}  
b'_0=b_f-\beta+2\ell_+=b_f+\ell \cos(\theta) \qquad -\ell_+<b_0<-\ell_-
\end{equation}
\begin{figure}
\includegraphics*[width=\Ws]{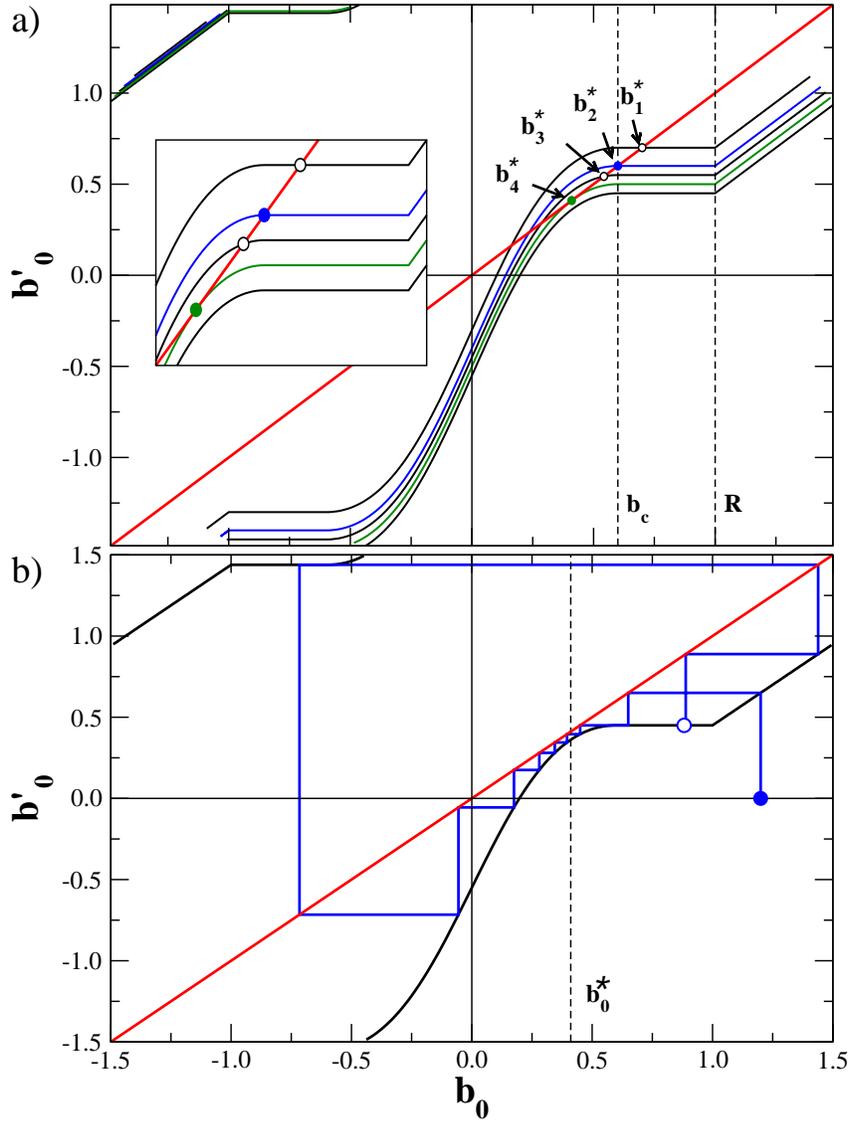}
\caption[]{a) Poincare map of the incoming collision parameter into itself for different forcing angles $\theta$. 
($\ell=2.5$; $f=2.0$; $b_m=\sqrt{3}/2$; $b_c=0.6$; $\beta_b=0.5$; $\theta_b=11.54^\circ$.)
The map has always the same form but is shifted down by a constant amount $\beta=\ell \sin(\theta)$.
The fixed points corresponding to increasing forcing angles are: $b^\star_1$: a fixed point in the region of irreversible collisions; 
$b^\star_2$: a fixed point that corresponds to successive collisions with $b_0=b_c$; 
$b^\star_3$: a fixed point corresponding to reversible collisions;
$b^\star_4$: a fixed point at the bifurcation angle $\theta=\theta_b$. 
The last map has no fixed points corresponding to $q=1$. 
b) A trajectory is shown for a forcing direction $\beta=0.55>\beta_b$. The trajectory has a periodicity $q=9$. 
Only one of the collisions leads to the particle crossing the top-bottom boundary 
of the unit cell and therefore the direction of the trajectory is [8,1].}
\label{mapF}
\end{figure}

 
We can then investigate the global trajectories by studying the above transformation of the impact parameter, which is 
in fact a Poincare map of the impact parameter into itself $b'_0=M(b_0)$ \cite{Arnold88}. 
Our objective is to compute the migration angle $\alpha$ as a function of the forcing angle $\theta$.
We showed before that for a given $\theta$ all the trajectories have the same migration angle. 
Therefore we can investigate single trajectories using the Poincare map to determine the asymptotic direction of motion.
This is particularly simple in the case of periodic trajectories.

\begin{figure}
\includegraphics*[width=\W]{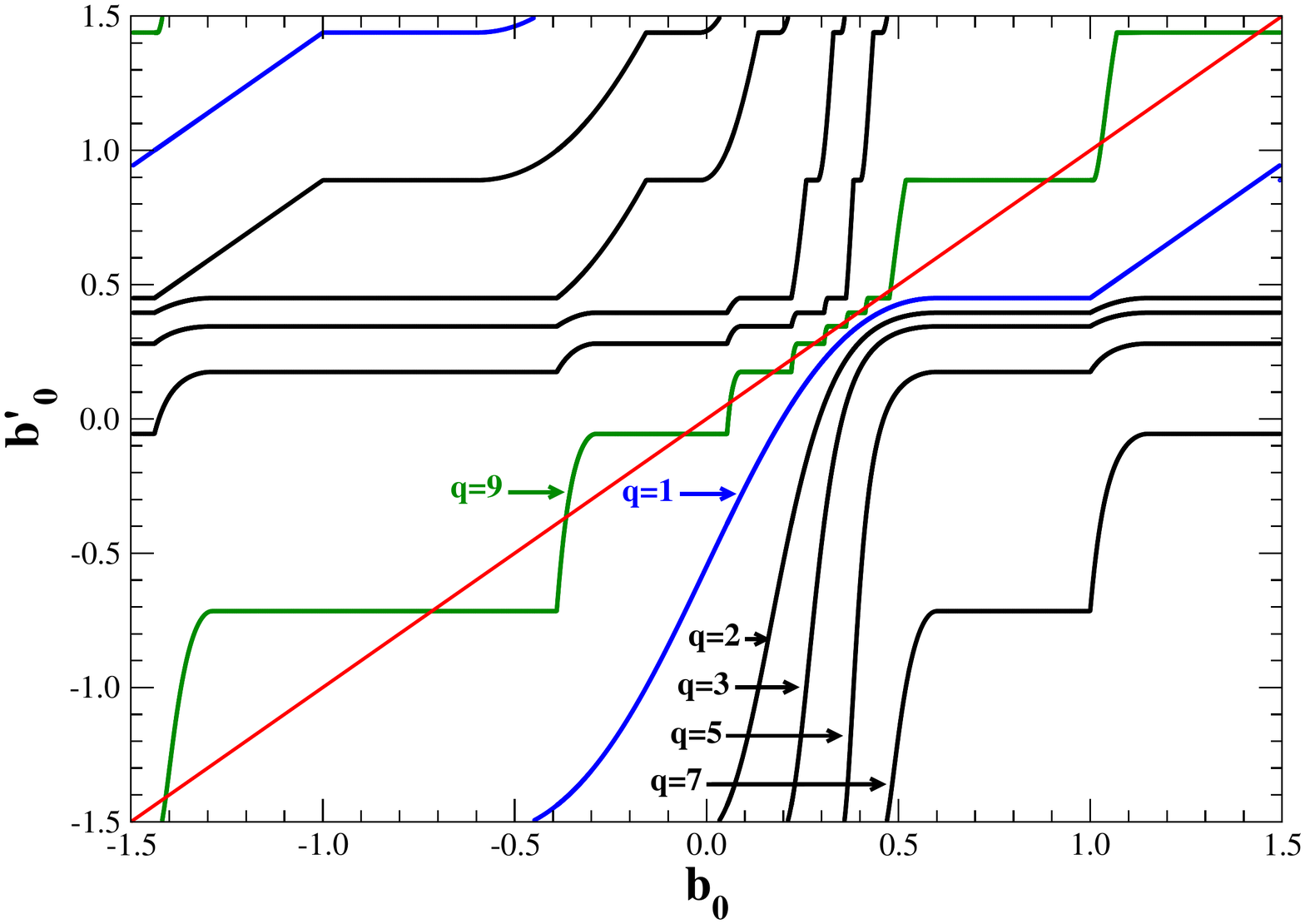}
\caption[]{ Poincare map of the incoming collision parameter into itself and compositions of the map of different 
orders $q=$1; 2; 3; 5; 7 and 9. The forcing angle is the same as in figure \ref{mapF}b, with $\beta=0.55>\beta_b$. 
($\ell=2.5$; $f=2.0$; $b_m=\sqrt{3}/2$; $b_c=0.6$; $\beta_b=0.5$; $\theta_b=11.54^\circ$.) It is clear that only
$q=9$ has fixed points, as shown in figure \ref{mapF}b by following an individual trajectory.
}
\label{map10}
\end{figure}

Let us define a periodic trajectory to have period $q$ if the particle goes through $q$ different collisions 
before repeating its motion (Note that all periodic trajectories are commensurate with the obstacle lattice) \cite{Ott02}. 
This corresponds to $q$ successive collisions with different incoming 
impact parameters before the sequence repeats itself. Periodic trajectories with period $q$ correspond to fixed 
points $b^\star_0$ of the $q$-times composed map of the impact parameter, for which $b^\star_0=M^{(q)}(b^\star_0)$. 
In figure \ref{mapF}a we present the map of the impact parameter into itself for increasing values of $\theta$. 
We also show the intersection points with the diagonal, which correspond to fixed points of period $q=1$ (the incoming
collision parameter is always the same). The presence of fixed points with $q=1$ indicates that the trajectory remains 
locked at $\alpha=0$ for sufficiently small forcing angles. 
In fact, for small forcing angles there is only one stable fixed point, located 
in the region of {\it irreversible} collisions (impact parameters between $b_c$ and $R$).
The fixed point corresponds to  $b^\star_0=1-\beta$ ($b^\star_1$ in the figure). 
The other fixed point is unstable, since the local slope of the map is greater than one \cite{Ott02}. 
As the forcing angle increases, the map is shifted down by an increasing amount, $\beta=\ell sin(\theta)$, 
and the fixed point moves to the left in the map, corresponding to smaller impact parameters.
Eventually, the fixed point reaches the critical impact parameter $b^\star_0=b_c$, as indicated in the figure ($b^\star_2$). 
For larger forcing angles, the stable fixed point is given by ($b^\star_3$ in the figure),
\begin{equation}
b^\star_0(\beta)=-\frac{\beta}{2}+\frac{1}{2f}\sqrt{2f^2-2+f^2\beta^2+(f^2-1)\sqrt{1-f^2\beta^2}}.
\end{equation}
The corresponding collision not only penetrates the parabolic regions but is also no longer {\it irreversible}.
However, these trajectories are still locked into the $\alpha=0$ overall motion.
Finally, as the forcing angle increases, the map goes through a tangent bifurcation for $\beta_b=\ell \sin(\theta_b)=1/f$ 
($b^\star_4$ in the figure), when
the stable and unstable fixed points meet at a single point $b^\star_b=b^\star_0(\beta_b)$, where the diagonal is tangent to the map 
(this bifurcation is sometimes also referred to as a {\em saddle node bifurcation} \cite{Eckmann82}).

Although in the vicinity of the bifurcation point the behavior is universal, the existence of {\it irreversible} 
collisions changes the global dynamics of the map, which in general is not chaotic. 
For example, for angles slightly larger than $\theta_b$, the map exhibits the universal {\it intermittent} 
behavior associated with tangent bifurcations, with long-lived intervals of quasi-periodic motion of period $q=1$, as the impact 
parameter goes across the $b_0 \sim b^\star_b$ region \cite{Eckmann82} (see ref. \cite{HuR82} for a discussion
of the corresponding renormalization-group approach in a generic tangent bifurcation, and ref. \cite{Baldovin06} for a more recent
perspective on tangent bifurcations in the context of Tsallis statistics).

However, the behavior outside this quasi-periodic region
is also periodic due to repeated {\it irreversible} collisions.  A typical trajectory with such intermittent behavior is shown in fig. \ref{mapF}b. The trajectory is quasi-periodic with $q=1$ for $b_0 \sim b^\star_b$. 
On the other hand, outside the near critical region, the trajectory is also periodic after two 
irreversible collisions, with period $q=9$ as shown in figures \ref{mapF}b and \ref{map10}. 

In general, there are two types of 
irreversible collisions, which we may call {\it positive} and {\it negative} ones, with $+1$ and $-1$ as the 
outgoing impact parameter, respectively, independent of the value of the incoming parameter. Therefore, a trajectory becomes periodic whenever
a second positive or negative collision occurs, which again highlights the commensurate nature of the periodic orbits.

\begin{figure}
\includegraphics*[width=\W]{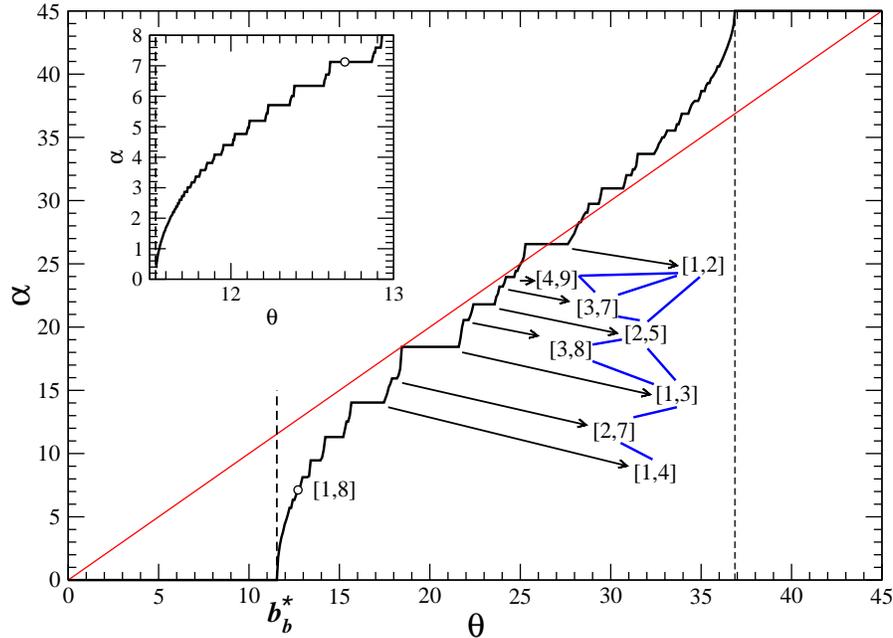}
\caption[]{Migration angle as a function of the forcing angle (same system as in figs. \ref{mapF} and \ref{map10}).
The critical forcing angle corresponding to a tangent bifurcation is shown with the dashed line. The inset shows the
bifurcation region in more detail. The open circle corresponds to the forcing angle $\theta=12.71^\circ$, 
$\beta=0.55>\beta_b$, $q=9$ and directional locking into the $[1,8]$ direction, also discussed in figs. \ref{mapF} 
and \ref{map10}}
\label{phaselock}
\end{figure}

We can still use the Poincare map to determine the phase locking behavior for all forcing angles. 
In the $q=9$ periodic trajectory shown in fig. \ref{mapF}b, for example, the particle goes through the 
bottom-top periodic boundary condition only once, which means that its asymptotic migration angle will
be $\arctan(1/8)$, as shown in figure \ref{phaselock}. Figure \ref{phaselock} also shows the behavior near the 
critical angle $\theta_c$ and the entire $\theta$ vs. $\alpha$ curve, which exhibits the typical
{\it Devil's staircase} structure \cite{Bak86}. We also show some of the observed locking angles
and their ordered structure, in which they form a Farey tree \cite{Schuster95} .

\subsection{Limiting Behavior}
\label{limitbehavior}
\begin{figure}
\includegraphics*[width=\W]{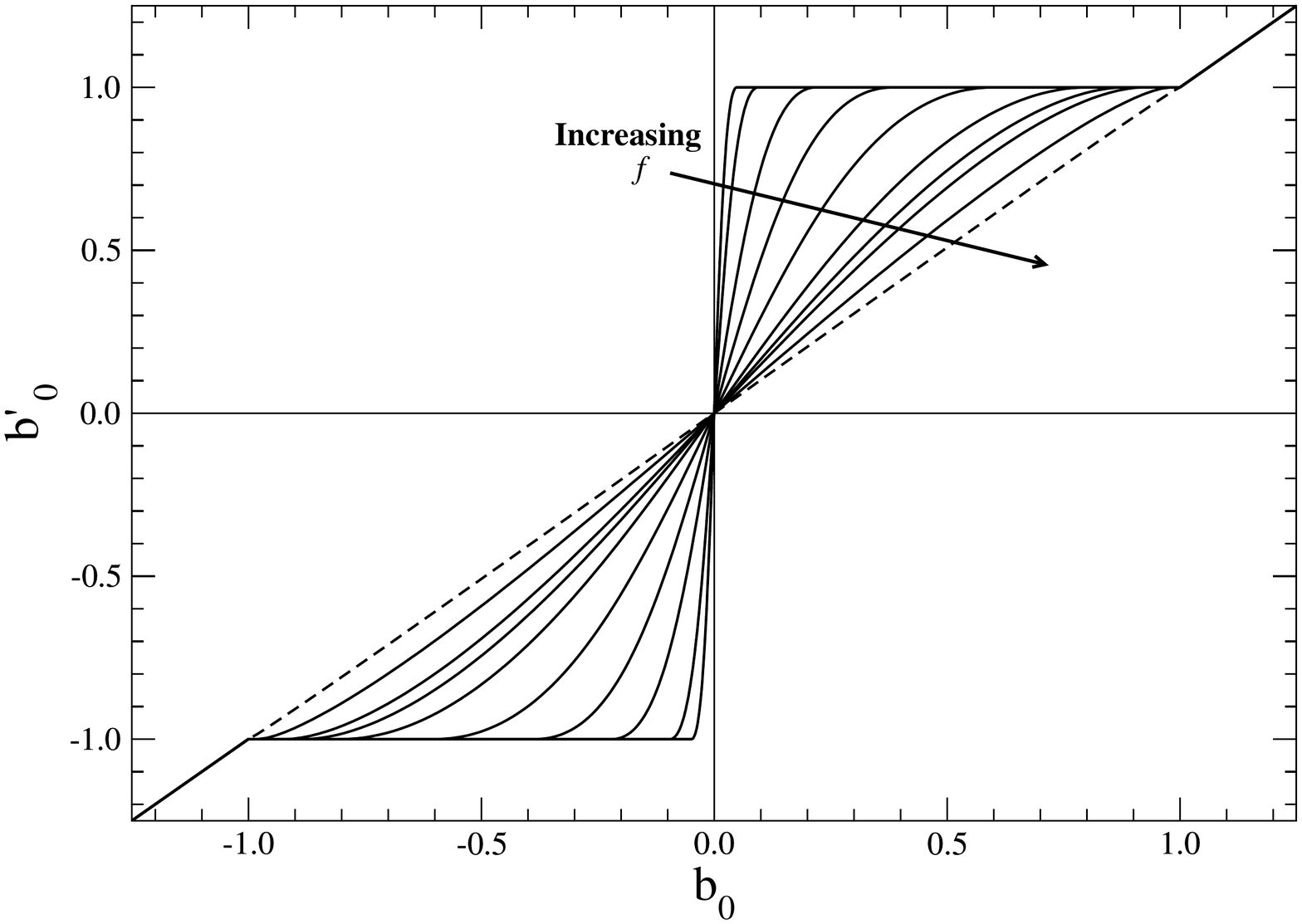}
\caption[]{Poincare map of the incoming impact parameter into itself for various relative magnitudes of the 
external force $f$. The dimensionless lattice parameter is $\ell=2.5$.}
\label{mapforce}
\end{figure}

In figure \ref{mapforce} we present the impact parameter map for different magnitudes of the external force.
In the limit of large external forces the repulsive centers have negligible influence on the trajectories,
and the map tends to an indentity relation, with both $b_c\to 1$ and $b_f \to b_0$. On the other hand, for
small forces, that is for forces $f \sim 1$, all collisions become irreversible, in that  
$b_c \to 0$ for $f \to 1$. Figure \ref{mapforce} shows that the map of the impact parameter tends to a 
piecewise-continuous map with only two regions: the region of unperturbed trajectories for $|b_0|>1$, and the region
of irreversible collisions for $|b_0|<1$. This is a particularly interesting limit, in that it corresponds to
impenetrable repulsive cores which could model, for example, the transport of a tracer particle through an
array of impermeable obstacles in the limit of high Peclet numbers, as we investigate in more detail in the 
next section. 

\section{Stochastic Transport: High Peclet number behavior}

In this section we investigate the stochastic transport of colloidal particles in the periodic landscape discussed in 
section \ref{subsec:potential}, {\it i. e.}, we consider diffusive transport in addition to the purely convective 
motion considered in the deterministic case. We will show that, at relatively high Peclet numbers, the average motion of
the particles exhibits directional locking equivalent to that observed in the deterministic case. 

\begin{figure}
\includegraphics*[width=\W]{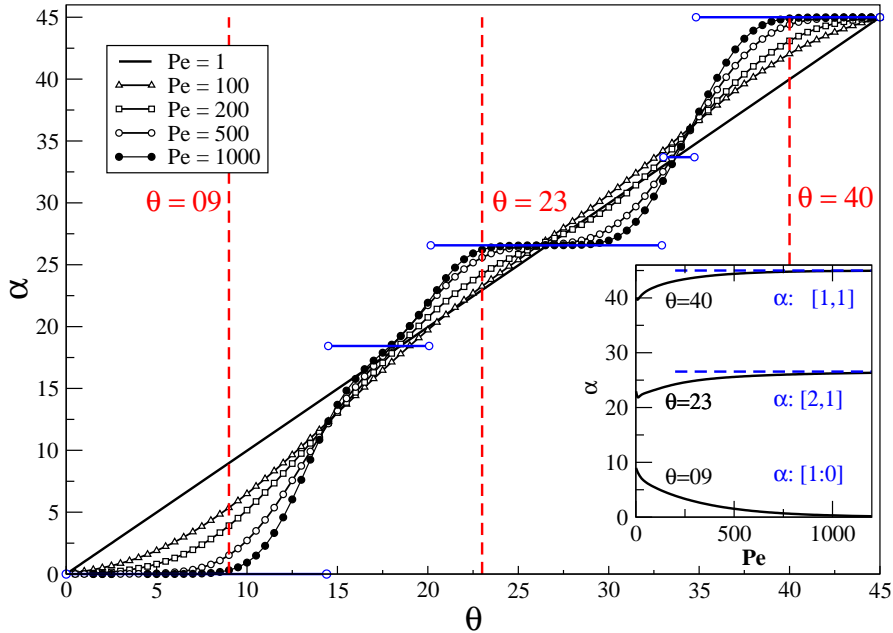}
\caption[]{Average migration angle calculated from Eq. \ref{avUx} for different forcing angles. 
The periodic landscape is described in section \ref{subsec:potential}. The dimensionless parameters are: $\ell=4.0$ and 
$f=1.0$. For all cases we computed the steady state probability distribution 
using a finite element method with over $3\times10^{5}$ degrees of freedom and
an element size smaller than $10^{-3}\times R$ close to the obstacle boundary.}
\label{comsol.soft.lock}
\end{figure}

In the presence of diffusive transport, the effective migration angle is given by the angle between the 
average velocity of the particles and the periodic lattice \cite{LiD07}. We first solve 
the Fokker-Planck equation for the probability density associated with the stochastic motion of the 
colloidal particles given by Eq. \ref{langevin}. The Fokker-Planck equation in non-dimensional 
variables reduces to:
\begin{equation}
\label{fp}
\frac{\partial}{\partial t}P(\mathbf{x},t)
+ f \frac{\partial}{\partial x}P(\mathbf{x},t)
-\frac{1}{\mathrm{Pe}} \nabla^2{P(\mathbf{x},t)}=0, \qquad r>1,
\end{equation}
where the Peclet number is given by $\mathrm{Pe}=F_{max}R/D\gamma$, with $D$ the diffusivity of the colloidal particle. 
The asymptotic distribution of colloidal particles in the unit cell, $P_\infty(\mathbf{x})$, corresponds
to the steady state solution of equation \ref{fp} above inside the unit cell and using periodic boundary 
conditions \cite{BrennerE93}. We can then obtain the components of the average migration velocity, $\langle U_i \rangle$, as well as the migration angle $\alpha$ from
their ratio \cite{BrennerE93},
\begin{equation}
\label{avUx}
\tan(\alpha)=
\frac{\langle U_y \rangle}{\langle U_x \rangle} = 
\left[\int_{0}^{\ell} dy ~ \left( - D \nabla P_\infty(\mathbf{x}) \right) \right] \bigg{/}
\left[\int_{0}^{\ell} dx ~ \left( f P_\infty(\mathbf{x}) - D \nabla P_\infty(\mathbf{x}) \right) \right].
\end{equation}

\begin{figure}
\includegraphics*[width=\Ws]{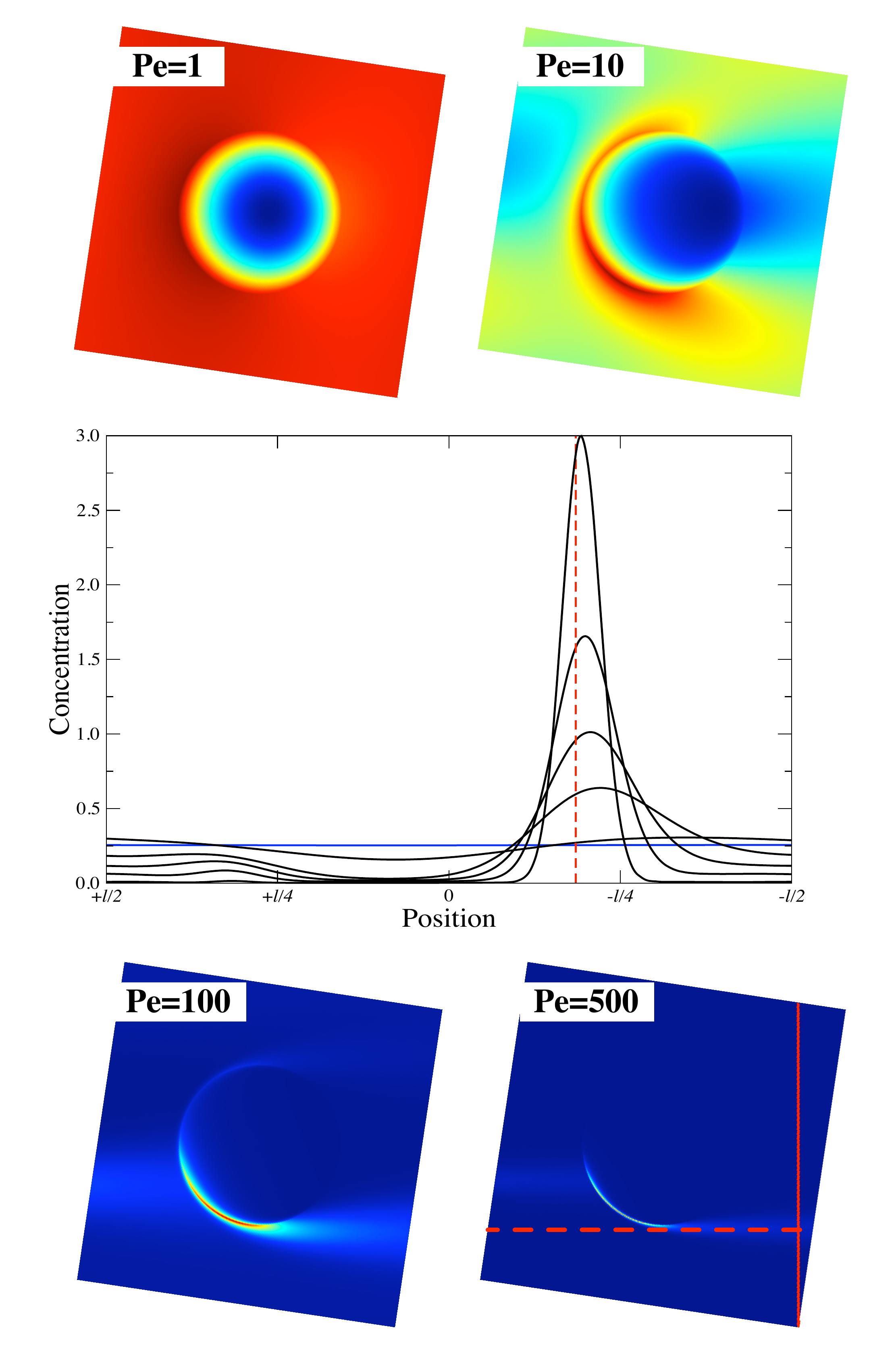}
\caption[]{Asymptotic distribution of particles in the unit cell for different Peclet numbers.
The dimensionless parameters are: $\ell=4.0$ and $f=1.0$. The forcing angle is $\theta=\arctan{0.15}\approx8.53^\circ$.
The plot at the center shows the concentration profile on a line along the $y$ axis (perpendicular to the external force). 
The dashed line in the plot marks the intersection with the tangent at the bottom of the circle. Both lines are shown
in the probability plot corresponding to Pe=500.}
\label{comsol}
\end{figure}

We obtained the stationary solution for the probability distribution using standard numerical methods.  
In figure \ref{comsol.soft.lock} we present the results for the average migration angle as a function of the
forcing angle for different Peclet numbers. It is clear that, as the Peclet number increases and the convective 
transport becomes dominant, the relation between the migration and the forcing angle tends to a structure
similar to those observed in the deterministic case, consisting of plateaus and steps. In fact, we show that the
average effective angle exhibits, in the limit of high Peclet numbers, the same directional locking as that 
predicted in the deterministic case (see figure \ref{comsol.soft.lock}).  

In figure \ref{comsol} we present the asymptotic probability distribution in the unit cell for different Peclet numbers.
We can see that, as the Peclet number increases and convective transport becomes dominant, the probability
flux is dominated by a stream that leaves the obstacles from the bottom of the circle and is parallel to
the force ($x$ axis). This is consistent with our interpretation in the deterministic case. Note that the solutions 
presented here correspond to the limiting case $f=1.0$ and, therefore, there is no penetration into the obstacles in the 
analogous deterministic case, as discussed in section \ref{limitbehavior}. 

The plot in figure \ref{comsol} shows 
the concentration profile on a line that is perpendicular to the forcing direction. It is clear that, as the Peclet 
number increases, the probability flux focuses on a narrow region that streams from the bottom point of the obstacle.
In fact, the probability maximum in the cross-section plot in figure \ref{comsol} tends to the point at which the
tangent to the circle parallel to the force intersects the cross-section line (indicated by a dashed line in the plot).

In the plots corresponding to Pe=100 and 500 we can see the {\it re-entrance} of the probability stream on the left boundary 
of the unit cell, as well as its {\it collision} with the obstacle. 
We can estimate the width of the probability peak to validate that it corresponds to diffusive spreading 
in the direction perpendicular to the convective motion. For this we follow the approximations done in the case of
geometric ratchets\cite{KellerMB02}, in which diffusion along the direction of the external force is neglected. 
In this case, 
\begin{equation}
\langle \Delta y^2 \rangle = 2 D \Delta t \sim 2 \left( \frac{\ell}{2U}\right) D = \frac{\ell}{\rm{Pe}},
\end{equation}
where $\ell/2$ is the distance between the streaming point and the cross-section measurement. This estimate agrees well with 
the numerical results. For example, for Pe=500 the previous Eq. predicts 
$\Delta y \sim 0.09$ and the width of the peak is $\sigma_y \sim 0.06$.

\begin{figure}
\includegraphics*[width=\W]{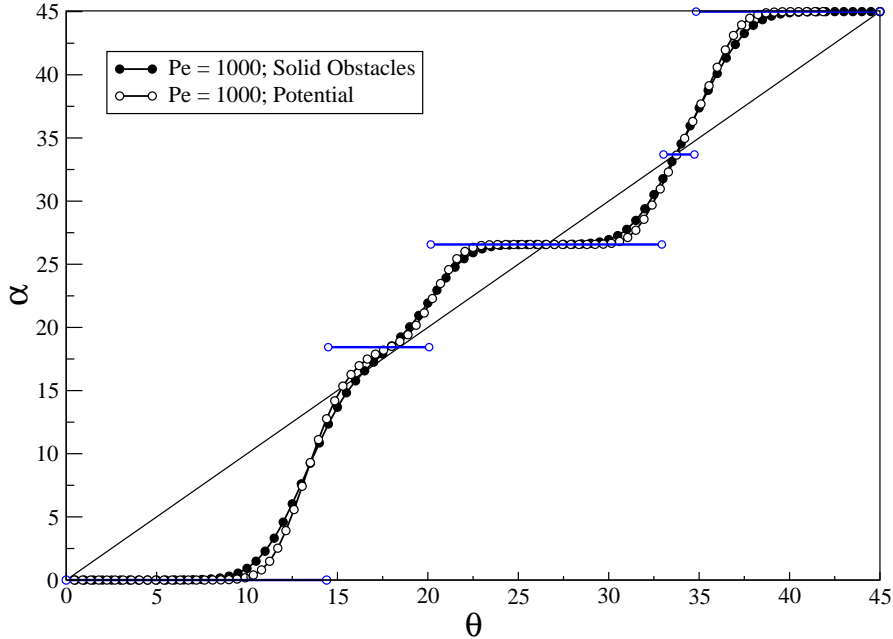}
\caption[]{Migration angle as a function of the forcing angle. Open symbols correspond to the parabolic repulsive centers with
non-dimensional lattice spacing  $\ell=4.0$, $f=1$, and Pe=1000. Solid symbols corresponds to numerical simulation in the
case of solid, non-permeable obstacles for the same lattice spacing and Peclet number.}
\label{hard}
\end{figure}

Finally, in figure \ref{hard} we compare the results of our model to the case of impermeable solid obstacles.
We discussed in the previous section that, in the limit of $f\sim1$ all collisions are irreversible ($b_c=0$ for $f=1$) 
and do not penetrate the obstacles. In this case we expect the dynamics to be similar to that in the case of solid obstacles.
This is only approximate, given that in the presence of diffusion, there is a non-zero probability of finding the particles
inside the repulsive regions. Figure \ref{hard} shows that, in fact, the migration angles obtained in our model are
very similar to the migration angles obtained in the case of solid obstacles for the same, relatively large, Peclet number.

\section{Relevance to Microfluidic devices}

In experimental work, it has been shown the possibility to separate particles 
using differences in the value of the bifurcation angle. In dimensional variables, the bifurcation
angle is given by $L \sin(\theta_b) = R/f$, which shows its dependence on the properties of the particles.
Specifically, the properties of the particles could come into play through the force ratio $f$ or through
the effective size of the repulsive centers $R$. 

In the case of optical lattices, both the
magnitude of the repulsive force as well as the characteristic size of the repulsive centers depend on the
particle size \cite{PeltonLG04}. In fact, our results for the bifurcation angle are analogous to the calculation
by \textcite{PeltonLG04} for the critical angle to escape a single barrier. Our results can then be used
to investigate the behavior at angles above the bifurcation.

Finally, we can investigate the transport of suspended particles in a pattern of solid obstacles using the limiting 
case $f\sim1$ discussed in \ref{limitbehavior}. 

In this case, $\sin(\theta_b)=R/L$, which depends on the size of the particles through
the effective size of the obstacles, $R=R_0+a$, where $R_0$ is the size of the obstacles and $a$ is the radius of
the particles. In this trivial approximation the bifurcation angle can also be obtained by straightforward geometrical 
considerations. The size dependency implies that larger particles will become unlocked from the $\alpha=0$ direction 
at larger angles of the driving force, as observed in experiments. 

Consider, for example, the separation of particles of two different sizes, $a_1=4\mu m$ and $a_2=6 \mu m$, 
in a lattice of (solid) cylindrical obstacles of radius $R_0=5\mu m$ and with a lattice constant $L=25\mu m$. 
Let us assume that the particles are driven at an average velocity of $U\approx10 \mu m/s$, 
which corresponds to a large value of the particles Peclet number, Pe$\sim10^3$. 
Then, considering only hard-sphere interactions (excluded volume effects) between the spheres and the obstacles 
the corresponding effective size of the obstacles is $R_1=9\mu m$ and $R_2=11 \mu m$ for particles of radius 
$a_1$ and $a_2$, respectively. Similarly, the dimensionless length depends on the size of the particles. Specifically,
$\ell_1=25/9$ and $\ell_2=25/11$, which results in different bifurcation angles for the two type of particles, 
$\theta_{b1}=21.1^\circ$ and $\theta_{b2}=26.1^\circ$ for sizes $a_1$ and $a_2$, respectively. Therefore, for a driving force
oriented at any angle $\theta$ relative to the obstacle lattice such that $\theta_{b1}<\theta<\theta_{b2}$, 
the particles will separate. In particular, the small particles will move in the lattice direction [1,2] with 
$\alpha_1=26.56^\circ$ and the large particles will be locked at $\alpha_2=0^\circ$.

\section{Summary}
We have shown that the transport of particles in a periodic lattice of repulsive centers exhibits analogous
behavior to that observed in microfluidic systems. The simplicity of the parabolic repulsive potentials 
allowed us to integrate the trajectories explicitly and showed the existence of periodic trajectories that
are commensurate with the obstacle lattice. We also showed that the motion can be determined by means of a
Poincare map of the incoming impact parameter into itself, which shows that there is a tangent bifurcation 
at the critical forcing angle for which the locking becomes different from $\alpha=0$. The entire migration-angle
vs. forcing-angle curve shows the characteristic Devil's staircase type of structure common to phase-locking systems.
Finally, we showed that the limiting behavior for impenetrable obstacles is equivalent to the high Peclet number limit
in the case of transport of particles in a periodic pattern of solid obstacles. Therefore, our previous results
provide insight into the separation problem in the case of periodic potential landscapes as well as in the case of 
periodic patterns of solid obstacles. In fact, we discuss a straightforward application of our results 
to calculating the bifurcation angle in both solid and optical lattices.

\section{Acknowledgements}
We thank R. Hansen for pointing to us the Farey-tree structure underlying the possible locking angles.
This material is partially based upon work supported by the National Science Foundation under 
Grant No. CBET-0731032.


\begin{thebibliography}{34}
\expandafter\ifx\csname natexlab\endcsname\relax\def\natexlab#1{#1}\fi
\expandafter\ifx\csname bibnamefont\endcsname\relax
  \def\bibnamefont#1{#1}\fi
\expandafter\ifx\csname bibfnamefont\endcsname\relax
  \def\bibfnamefont#1{#1}\fi
\expandafter\ifx\csname citenamefont\endcsname\relax
  \def\citenamefont#1{#1}\fi
\expandafter\ifx\csname url\endcsname\relax
  \def\url#1{\texttt{#1}}\fi
\expandafter\ifx\csname urlprefix\endcsname\relax\def\urlprefix{URL }\fi
\providecommand{\bibinfo}[2]{#2}
\providecommand{\eprint}[2][]{\url{#2}}

\bibitem[{\citenamefont{Dorfman and Brenner}(2001)}]{DorfmanB01}
\bibinfo{author}{\bibfnamefont{K.~D.} \bibnamefont{Dorfman}} \bibnamefont{and}
  \bibinfo{author}{\bibfnamefont{H.}~\bibnamefont{Brenner}},
  \bibinfo{journal}{J. Colloid Interface Sci.} \textbf{\bibinfo{volume}{238}},
  \bibinfo{pages}{390} (\bibinfo{year}{2001}), ISSN \bibinfo{issn}{0021-9797}.

\bibitem[{\citenamefont{Dorfman and Brenner}(2002)}]{DorfmanB02}
\bibinfo{author}{\bibfnamefont{K.~D.} \bibnamefont{Dorfman}} \bibnamefont{and}
  \bibinfo{author}{\bibfnamefont{H.}~\bibnamefont{Brenner}},
  \bibinfo{journal}{Phys. Rev. E} \textbf{\bibinfo{volume}{65}},
  \bibinfo{pages}{052103} (\bibinfo{year}{2002}), ISSN
  \bibinfo{issn}{1063-651X}.

\bibitem[{\citenamefont{Huang et~al.}(2004)\citenamefont{Huang, Cox, Austin,
  and Sturm}}]{HuangCAS04}
\bibinfo{author}{\bibfnamefont{L.~R.} \bibnamefont{Huang}},
  \bibinfo{author}{\bibfnamefont{E.~C.} \bibnamefont{Cox}},
  \bibinfo{author}{\bibfnamefont{R.~H.} \bibnamefont{Austin}},
  \bibnamefont{and} \bibinfo{author}{\bibfnamefont{J.~C.} \bibnamefont{Sturm}},
  \bibinfo{journal}{Science} \textbf{\bibinfo{volume}{304}},
  \bibinfo{pages}{987} (\bibinfo{year}{2004}),
  \urlprefix\url{http://www.sciencemag.org/cgi/content/abstract/304/5673/987}.

\bibitem[{\citenamefont{Beech and Tegenfeldt}(2008)}]{BeechBT08}
\bibinfo{author}{\bibfnamefont{J.~P.} \bibnamefont{Beech}} \bibnamefont{and}
  \bibinfo{author}{\bibfnamefont{J.~O.} \bibnamefont{Tegenfeldt}},
  \bibinfo{journal}{Lab chip} \textbf{\bibinfo{volume}{8}},
  \bibinfo{pages}{657} (\bibinfo{year}{2008}).

\bibitem[{\citenamefont{Morton et~al.}(2008{\natexlab{a}})\citenamefont{Morton,
  Loutherback, Inglis, Tsui, Sturm, Chou, and Austin}}]{MortonLITSCA08}
\bibinfo{author}{\bibfnamefont{K.~J.} \bibnamefont{Morton}},
  \bibinfo{author}{\bibfnamefont{K.}~\bibnamefont{Loutherback}},
  \bibinfo{author}{\bibfnamefont{D.~W.} \bibnamefont{Inglis}},
  \bibinfo{author}{\bibfnamefont{O.~K.} \bibnamefont{Tsui}},
  \bibinfo{author}{\bibfnamefont{J.~C.} \bibnamefont{Sturm}},
  \bibinfo{author}{\bibfnamefont{S.~Y.} \bibnamefont{Chou}}, \bibnamefont{and}
  \bibinfo{author}{\bibfnamefont{R.~H.} \bibnamefont{Austin}},
  \bibinfo{journal}{Proc. Natl. Acad. Sci.} pp. \bibinfo{pages}{7434--7438}
  (\bibinfo{year}{2008}{\natexlab{a}}),
  \urlprefix\url{http://www.pnas.org/cgi/content/abstract/0712398105v1}.

\bibitem[{\citenamefont{Morton et~al.}(2008{\natexlab{b}})\citenamefont{Morton,
  Loutherback, Inglis, Tsui, Sturm, Chou, and Austin}}]{MortonLITSCA08b}
\bibinfo{author}{\bibfnamefont{K.~J.} \bibnamefont{Morton}},
  \bibinfo{author}{\bibfnamefont{K.}~\bibnamefont{Loutherback}},
  \bibinfo{author}{\bibfnamefont{D.~W.} \bibnamefont{Inglis}},
  \bibinfo{author}{\bibfnamefont{O.~K.} \bibnamefont{Tsui}},
  \bibinfo{author}{\bibfnamefont{J.~C.} \bibnamefont{Sturm}},
  \bibinfo{author}{\bibfnamefont{S.~Y.} \bibnamefont{Chou}}, \bibnamefont{and}
  \bibinfo{author}{\bibfnamefont{R.~H.} \bibnamefont{Austin}},
  \bibinfo{journal}{Lab chip} \textbf{\bibinfo{volume}{8}},
  \bibinfo{pages}{1448} (\bibinfo{year}{2008}{\natexlab{b}}),
  \urlprefix\url{http://dx.doi.org/10.1039/b805614e}.

\bibitem[{\citenamefont{Korda et~al.}(2002)\citenamefont{Korda, Taylor, and
  Grier}}]{KordaTG02}
\bibinfo{author}{\bibfnamefont{P.~T.} \bibnamefont{Korda}},
  \bibinfo{author}{\bibfnamefont{M.~B.} \bibnamefont{Taylor}},
  \bibnamefont{and} \bibinfo{author}{\bibfnamefont{D.~G.} \bibnamefont{Grier}},
  \bibinfo{journal}{Phys. Rev. Lett.} \textbf{\bibinfo{volume}{89}},
  \bibinfo{pages}{128301} (\bibinfo{year}{2002}), ISSN
  \bibinfo{issn}{0031-9007}.

\bibitem[{\citenamefont{Ladavac et~al.}(2004)\citenamefont{Ladavac, Kasza, and
  Grier}}]{LadavacLKG04}
\bibinfo{author}{\bibfnamefont{K.}~\bibnamefont{Ladavac}},
  \bibinfo{author}{\bibfnamefont{K.}~\bibnamefont{Kasza}}, \bibnamefont{and}
  \bibinfo{author}{\bibfnamefont{D.}~\bibnamefont{Grier}},
  \bibinfo{journal}{Phys. Rev. E} \textbf{\bibinfo{volume}{70}}
  (\bibinfo{year}{2004}).

\bibitem[{\citenamefont{Gopinathan and Grier}(2004)}]{AjayG04}
\bibinfo{author}{\bibfnamefont{A.}~\bibnamefont{Gopinathan}} \bibnamefont{and}
  \bibinfo{author}{\bibfnamefont{D.~G.} \bibnamefont{Grier}},
  \bibinfo{journal}{Phys. Rev. Lett.} \textbf{\bibinfo{volume}{92}},
  \bibinfo{pages}{130602} (\bibinfo{year}{2004}).

\bibitem[{\citenamefont{Roichman et~al.}(2007)\citenamefont{Roichman, Wong, and
  Grier}}]{RoichmanRWG07}
\bibinfo{author}{\bibfnamefont{Y.}~\bibnamefont{Roichman}},
  \bibinfo{author}{\bibfnamefont{V.}~\bibnamefont{Wong}}, \bibnamefont{and}
  \bibinfo{author}{\bibfnamefont{D.~G.} \bibnamefont{Grier}},
  \bibinfo{journal}{Phys. Rev. E} \textbf{\bibinfo{volume}{75}}
  (\bibinfo{year}{2007}).

\bibitem[{\citenamefont{Pelton et~al.}(2004)\citenamefont{Pelton, Ladavac, and
  Grier}}]{PeltonLG04}
\bibinfo{author}{\bibfnamefont{M.}~\bibnamefont{Pelton}},
  \bibinfo{author}{\bibfnamefont{K.}~\bibnamefont{Ladavac}}, \bibnamefont{and}
  \bibinfo{author}{\bibfnamefont{D.}~\bibnamefont{Grier}},
  \bibinfo{journal}{Phys. Rev. E} \textbf{\bibinfo{volume}{70}}
  (\bibinfo{year}{2004}).

\bibitem[{\citenamefont{Reichhardt et~al.}(2002)\citenamefont{Reichhardt,
  Olson, and Hastings}}]{ReichhardtOH02}
\bibinfo{author}{\bibfnamefont{C.}~\bibnamefont{Reichhardt}},
  \bibinfo{author}{\bibfnamefont{C.~J.} \bibnamefont{Olson}}, \bibnamefont{and}
  \bibinfo{author}{\bibfnamefont{M.~B.} \bibnamefont{Hastings}},
  \bibinfo{journal}{Phys. Rev. Lett.} \textbf{\bibinfo{volume}{89}}
  (\bibinfo{year}{2002}), ISSN \bibinfo{issn}{0031-9007}.

\bibitem[{\citenamefont{Reichhardt and Nori}(1999)}]{ReichhardtF99}
\bibinfo{author}{\bibfnamefont{C.}~\bibnamefont{Reichhardt}} \bibnamefont{and}
  \bibinfo{author}{\bibfnamefont{F.}~\bibnamefont{Nori}},
  \bibinfo{journal}{Phys. Rev. Lett.} \textbf{\bibinfo{volume}{82}},
  \bibinfo{pages}{414} (\bibinfo{year}{1999}), ISSN \bibinfo{issn}{0031-9007}.

\bibitem[{\citenamefont{Marconi et~al.}(2000)\citenamefont{Marconi, Candia,
  Balenzuela, Pastoriza, Dom{\'\i}nguez, and Martinoli}}]{MarconiCBPDM00}
\bibinfo{author}{\bibfnamefont{V.~I.} \bibnamefont{Marconi}},
  \bibinfo{author}{\bibfnamefont{S.}~\bibnamefont{Candia}},
  \bibinfo{author}{\bibfnamefont{P.}~\bibnamefont{Balenzuela}},
  \bibinfo{author}{\bibfnamefont{H.}~\bibnamefont{Pastoriza}},
  \bibinfo{author}{\bibfnamefont{D.}~\bibnamefont{Dom{\'\i}nguez}},
  \bibnamefont{and}
  \bibinfo{author}{\bibfnamefont{P.}~\bibnamefont{Martinoli}},
  \bibinfo{journal}{Phys. Rev. B} \textbf{\bibinfo{volume}{62}},
  \bibinfo{pages}{4096} (\bibinfo{year}{2000}),
  \urlprefix\url{http://link.aps.org/abstract/PRB/v62/p4096}.

\bibitem[{\citenamefont{Reichhardt et~al.}(2004)\citenamefont{Reichhardt,
  Reichhardt, and Hastings}}]{ReichhardtRH04}
\bibinfo{author}{\bibfnamefont{C.}~\bibnamefont{Reichhardt}},
  \bibinfo{author}{\bibfnamefont{C.~J.~O.} \bibnamefont{Reichhardt}},
  \bibnamefont{and} \bibinfo{author}{\bibfnamefont{M.~B.}
  \bibnamefont{Hastings}}, \bibinfo{journal}{Phys. Rev. E}
  \textbf{\bibinfo{volume}{69}}, \bibinfo{pages}{056115}
  (\bibinfo{year}{2004}), ISSN \bibinfo{issn}{1063-651X}.

\bibitem[{\citenamefont{Reichhardt and Reichhardt}(2004)}]{ReichhardtR04}
\bibinfo{author}{\bibfnamefont{C.}~\bibnamefont{Reichhardt}} \bibnamefont{and}
  \bibinfo{author}{\bibfnamefont{C.~J.~O.} \bibnamefont{Reichhardt}},
  \bibinfo{journal}{Phys. Rev. E} \textbf{\bibinfo{volume}{69}},
  \bibinfo{pages}{041405} (\bibinfo{year}{2004}), ISSN
  \bibinfo{issn}{1063-651X}.

\bibitem[{\citenamefont{Ott}(2002)}]{Ott02}
\bibinfo{author}{\bibfnamefont{E.}~\bibnamefont{Ott}},
  \emph{\bibinfo{title}{Chaos in Dynamical Systems}}
  (\bibinfo{publisher}{Cambridge University Press}, \bibinfo{year}{2002}),
  \bibinfo{edition}{2nd} ed., ISBN \bibinfo{isbn}{9780521010849}.

\bibitem[{\citenamefont{Strogatz}(1994)}]{Strogatz94}
\bibinfo{author}{\bibfnamefont{S.~H.} \bibnamefont{Strogatz}},
  \emph{\bibinfo{title}{Nonlinear dynamics and Chaos: with applications to
  physics, biology, chemistry, and engineering}}
  (\bibinfo{publisher}{Addison-Wesley Pub.}, \bibinfo{address}{Reading, Mass.},
  \bibinfo{year}{1994}), ISBN \bibinfo{isbn}{0201543443},
  \urlprefix\url{http://www.loc.gov/catdir/enhancements/fy0830/93006166-d.html%
}.

\bibitem[{\citenamefont{Lacasta et~al.}(2005)\citenamefont{Lacasta, Sancho,
  Romero, and Lindenberg}}]{LacastaSRL05}
\bibinfo{author}{\bibfnamefont{A.}~\bibnamefont{Lacasta}},
  \bibinfo{author}{\bibfnamefont{J.}~\bibnamefont{Sancho}},
  \bibinfo{author}{\bibfnamefont{A.}~\bibnamefont{Romero}}, \bibnamefont{and}
  \bibinfo{author}{\bibfnamefont{K.}~\bibnamefont{Lindenberg}},
  \bibinfo{journal}{Phys. Rev. Lett.} \textbf{\bibinfo{volume}{94}},
  \bibinfo{pages}{160601} (\bibinfo{year}{2005}).

\bibitem[{\citenamefont{Sancho et~al.}(2005)\citenamefont{Sancho, Khoury,
  Lindenberg, and Lacasta}}]{SanchoKLL05}
\bibinfo{author}{\bibfnamefont{J.~M.} \bibnamefont{Sancho}},
  \bibinfo{author}{\bibfnamefont{M.}~\bibnamefont{Khoury}},
  \bibinfo{author}{\bibfnamefont{K.}~\bibnamefont{Lindenberg}},
  \bibnamefont{and} \bibinfo{author}{\bibfnamefont{A.~M.}
  \bibnamefont{Lacasta}}, \bibinfo{journal}{J. Phys. Condens. Matter}
  \textbf{\bibinfo{volume}{17}}, \bibinfo{pages}{S4151} (\bibinfo{year}{2005}),
  ISSN \bibinfo{issn}{0953-8984}.

\bibitem[{\citenamefont{Lacasta et~al.}(2006)\citenamefont{Lacasta, Khoury,
  Sancho, and Lindenberg}}]{LacastaKSL06}
\bibinfo{author}{\bibfnamefont{A.~M.} \bibnamefont{Lacasta}},
  \bibinfo{author}{\bibfnamefont{M.}~\bibnamefont{Khoury}},
  \bibinfo{author}{\bibfnamefont{J.~M.} \bibnamefont{Sancho}},
  \bibnamefont{and}
  \bibinfo{author}{\bibfnamefont{K.}~\bibnamefont{Lindenberg}},
  \bibinfo{journal}{Mod. Phys. Lett. B} \textbf{\bibinfo{volume}{20}},
  \bibinfo{pages}{1427} (\bibinfo{year}{2006}), ISSN \bibinfo{issn}{0217-9849}.

\bibitem[{\citenamefont{Guantes and Miret-Art{\'e}s}(2003)}]{GuantesM03}
\bibinfo{author}{\bibfnamefont{R.}~\bibnamefont{Guantes}} \bibnamefont{and}
  \bibinfo{author}{\bibfnamefont{S.}~\bibnamefont{Miret-Art{\'e}s}},
  \bibinfo{journal}{Phys. Rev. E} \textbf{\bibinfo{volume}{67}}
  (\bibinfo{year}{2003}),
  \urlprefix\url{http://link.aps.org/abstract/PRE/v67/e046212}.

\bibitem[{\citenamefont{Gleeson et~al.}(2006)\citenamefont{Gleeson, Sancho,
  Lacasta, and Lindenberg}}]{GleesonSLL06}
\bibinfo{author}{\bibfnamefont{J.~P.} \bibnamefont{Gleeson}},
  \bibinfo{author}{\bibfnamefont{J.~M.} \bibnamefont{Sancho}},
  \bibinfo{author}{\bibfnamefont{A.~M.} \bibnamefont{Lacasta}},
  \bibnamefont{and}
  \bibinfo{author}{\bibfnamefont{K.}~\bibnamefont{Lindenberg}},
  \bibinfo{journal}{Phys. Rev. E} \textbf{\bibinfo{volume}{73}},
  \bibinfo{pages}{041102} (\bibinfo{year}{2006}), ISSN
  \bibinfo{issn}{1539-3755}.

\bibitem[{\citenamefont{{Mc Quarrie}}(2000)}]{McQuarrie00}
\bibinfo{author}{\bibfnamefont{D.~A.} \bibnamefont{{Mc Quarrie}}},
  \emph{\bibinfo{title}{Statistical Mechanics}} (\bibinfo{publisher}{University
  Science Books}, \bibinfo{year}{2000}), \bibinfo{edition}{2nd} ed., ISBN
  \bibinfo{isbn}{9781891389153}.

\bibitem[{\citenamefont{Hannes~Risken}(1996)}]{Risken}
\bibinfo{author}{\bibfnamefont{T.~F.} \bibnamefont{Hannes~Risken}},
  \emph{\bibinfo{title}{The Fokker-Planck Equation: Methods of Solutions and
  Applications}} (\bibinfo{publisher}{Springer}, \bibinfo{year}{1996}),
  \bibinfo{edition}{2nd} ed., ISBN \bibinfo{isbn}{9783540615309}.

\bibitem[{\citenamefont{Arnold}(1996)}]{Arnold88}
\bibinfo{author}{\bibfnamefont{V.~I.} \bibnamefont{Arnold}},
  \emph{\bibinfo{title}{Geometrical Methods in the Theory of Ordinary
  Differential Equations}} (\bibinfo{publisher}{Springer},
  \bibinfo{year}{1996}), \bibinfo{edition}{2nd} ed., ISBN
  \bibinfo{isbn}{9780387966496}.

\bibitem[{\citenamefont{Eckmann}(1982)}]{Eckmann82}
\bibinfo{author}{\bibfnamefont{J.~P.} \bibnamefont{Eckmann}},
  \bibinfo{journal}{Rev. Mod. Phys.} \textbf{\bibinfo{volume}{53}},
  \bibinfo{pages}{643} (\bibinfo{year}{1982}).

\bibitem[{\citenamefont{Hu and Rudnick}(1982)}]{HuR82}
\bibinfo{author}{\bibfnamefont{B.}~\bibnamefont{Hu}} \bibnamefont{and}
  \bibinfo{author}{\bibfnamefont{J.}~\bibnamefont{Rudnick}},
  \bibinfo{journal}{Phys. Rev. Lett.} \textbf{\bibinfo{volume}{48}},
  \bibinfo{pages}{1645} (\bibinfo{year}{1982}).

\bibitem[{\citenamefont{Baldovin}(2006)}]{Baldovin06}
\bibinfo{author}{\bibfnamefont{F.}~\bibnamefont{Baldovin}},
  \bibinfo{journal}{Phys. A} \textbf{\bibinfo{volume}{372}},
  \bibinfo{pages}{224} (\bibinfo{year}{2006}).

\bibitem[{\citenamefont{Bak}(1986)}]{Bak86}
\bibinfo{author}{\bibfnamefont{P.}~\bibnamefont{Bak}}, \bibinfo{journal}{Phys.
  Today} \textbf{\bibinfo{volume}{39}}, \bibinfo{pages}{38}
  (\bibinfo{year}{1986}).

\bibitem[{\citenamefont{Schuster}(1995)}]{Schuster95}
\bibinfo{author}{\bibfnamefont{H.~G.} \bibnamefont{Schuster}},
  \emph{\bibinfo{title}{Deterministic Chaos}} (\bibinfo{publisher}{VCH},
  \bibinfo{address}{New York}, \bibinfo{year}{1995}).

\bibitem[{\citenamefont{Li and Drazer}(2007)}]{LiD07}
\bibinfo{author}{\bibfnamefont{Z.}~\bibnamefont{Li}} \bibnamefont{and}
  \bibinfo{author}{\bibfnamefont{G.}~\bibnamefont{Drazer}},
  \bibinfo{journal}{Phys. Rev. Lett.} \textbf{\bibinfo{volume}{98}},
  \bibinfo{pages}{050602} (\bibinfo{year}{2007}), ISSN
  \bibinfo{issn}{0031-9007}.

\bibitem[{\citenamefont{David~Edwards}(1993)}]{BrennerE93}
\bibinfo{author}{\bibfnamefont{H.~B.} \bibnamefont{David~Edwards}},
  \emph{\bibinfo{title}{Macrotransport Processes}}
  (\bibinfo{publisher}{Butterworth-Heinemann}, \bibinfo{year}{1993}), ISBN
  \bibinfo{isbn}{9780750693325}.

\bibitem[{\citenamefont{Keller et~al.}(2002)\citenamefont{Keller, Marquardt,
  and Bruder}}]{KellerMB02}
\bibinfo{author}{\bibfnamefont{C.}~\bibnamefont{Keller}},
  \bibinfo{author}{\bibfnamefont{F.}~\bibnamefont{Marquardt}},
  \bibnamefont{and} \bibinfo{author}{\bibfnamefont{C.}~\bibnamefont{Bruder}},
  \bibinfo{journal}{Phys. Rev. E} \textbf{\bibinfo{volume}{65}},
  \bibinfo{pages}{041927} (\bibinfo{year}{2002}), ISSN
  \bibinfo{issn}{1063-651X}.

\end{thebibliography}
\end{document}